\documentclass[11pt,twoside]{article}

\usepackage{asp2006}
\usepackage{epsf}
\usepackage{psfig}
\usepackage{lscape}
\usepackage{graphicx}

\begin{document}

\title{Simulations of the Magneto-rotational Instability in
  Core-Collapse Supernovae}

\author{M.~Obergaulinger, P.~Cerd{\'a}-Dur{\'a}n, and E.~M{\"u}ller}

\affil{Max-Planck-Institut f\"ur Astrophysik, Postfach 1317, D-85741
  Garching, Germany}

\author{M.A.~Aloy}

\affil{Departament d{'{}}Astronomia i Astrof{\'i}sica, Universitat de
  Val{\`e}ncia, Edifici d{'{}}Investigaci{\'o} Jeroni Mu{\~n}oz, C/
  Dr. Moliner, 50, E-46100 Burjassot (Val{\`e}ncia), Spain }

\begin{abstract}
  We assess the importance of the magneto-rotational instability in
  core-collapse supernovae by an analysis of the growth rates of
  unstable modes in typical post-collapse systems and by numerical
  simulations of simplified models.  The interplay of differential
  rotation and thermal stratification defines different instability
  regimes which we confirm in our simulations.  We investigate the
  termination of the growth of the MRI by parasitic instabilities,
  establish scaling laws characterising the termination amplitude, and
  study the long-term evolution of the saturated turbulent state.
\end{abstract}

\section{Introduction}
\label{Sek:Intro}

At the end of its hydrostatic evolution, the inner core of a star of
more than about eight solar masses, consisting of about 1.5 solar
masses of iron or oxygen, neon, and magnesium loses support against
its self-gravity and collapses to supranuclear density
$\rho_{\mathrm{nuc}} \geq 2 \times 10^{14} ~ \mathrm{g\,cm^{-3}}$,
releasing a gravitational binding energy $E_{\mathrm{grv}} \sim
10^{53} ~ \mathrm{erg}$.  When this density is reached, a shock wave
is launched and starts to propagate outward.  Still within the core,
the shock wave stalls and turns into a stationary accretion shock as
energy is lost when nuclei are dissociated in the post-shock region:
the prompt explosion fails.

Most of the gravitational binding energy liberated during collapse
leaves the star in form of neutrinos, but a small ($\sim 1\%$)
fraction is used to unbind the envelope and power a supernova (SN)
explosion.  How a fraction of $E_{\mathrm{grv}}$ is transferred to the
shock wave and the surrounding envelope is still not fully understood,
and forms the central issue of SN theory.

This is not a problem of energy budget but of energy transfer.  The
gravitational binding energy of the envelope ($\sim 10^{50} ~
\mathrm{erg}$) as well as the kinetic and electromagnetic energies of
a SN ($\sim 10^{49}$ and $\sim 10^{51} ~ \mathrm{erg}$, respectively)
are much smaller than $E_{\mathrm{grv}}$.  Consequently, the transfer
of a small fraction of $E_{\mathrm{grv}}$ suffices to revive the shock
wave and to trigger the explosion.  However, it is extremely difficult
to tap efficiently the energy carried by weakly interacting neutrinos.
The mechanisms proposed to account for shock revival and explosion can
be grouped into the following categories:
\begin{description}
\item[The spherical neutrino mechanism] Heating of the post-shock
  matter by interactions with the neutrinos diffusing out of the core
  revives the shock \citep[e.g.,][]{Bethe__1990__RvMP__SN_mech}.  This
  mechanism works for stars around $8 ~ \mathrm{M}_{\odot}$ due to a
  steep density profile facilitating shock propagation
  \citep{Kitaura_Janka_Hillebrandt__2006__aap__Explosions_of_O-Ne-Mg_cores_theCrab_supernova_and_subluminous_typeII-P_supernovae}.
\item[Hydrodynamic instabilities] Without the restriction to spherical
  symmetry, hydrodynamic instabilities can develop: convection in the
  proto-neutron star (PNS) and behind the shock and the standing
  accretion shock instability (SASI) can enhance the efficiency of
  neutrino heating and the transfer of binding energy of the accreting
  gas to the surrounding matter
  \citep[][]{Blondin_Mezzacappa_DeMarino__2003__apj__Stability_of_Standing_Accretion_Shocks_with_an_Eye_toward_Core-Collapse_Supernovae,Foglizzo_et_al__2007__apj__Instability_of_a_Stalled_Accretion_Shock_AAC}.
  They may also account for the pronounced asphericities of the
  explosion and the observed pulsar kicks
  \citep[e.g.,][]{Scheck_et_al__2006__AA__2d_Aniso}.  While this
  mechanism has proven successful in detailed 2d simulations of stars
  up to 15 solar masses employing state-of-the-art microphysics
  \citep{Marek_etal__2009__AA__EOS-depnc_nu_GW}, its robustness and
  applicability to stars of higher mass is still uncertain.
\item[Energy transfer by waves] Apart from neutrinos, waves may carry
  energy from the central region to the surroundings.  Unstable
  g-modes in the PNS excited by accreting matter can lead to the
  emission of acoustic waves travelling upwards and depositing energy
  near the shock
  \citep[][]{Burrows_et_al__2006__apj__A_New_Mechanism_for_Core-Collapse_Supernova_Explosions}.
  A similar effect can be caused by Alfv\'en waves excited in the
  convective region of a magnetised PNS
  \citep[][]{Suzuki_Sumiyoshi_Yamada__2008__ApJ__Alfven_driven_SN}.
  The importance of acoustic waves has been suggested in simulations,
  whereas the heating by Alfv\'en waves has not been explored
  thoroughly.
\item[Rotational mechanisms] Any rotation of the progenitor will be
  amplified during the collapse; additionally, differential rotation
  will be created even for an initially rigidly rotating progenitor.
  If sufficiently rapid, rotation may provide sufficient additional
  energy to unbind the post-shock matter.  In contrast to the energy
  of neutrinos, rotational energy can be tapped fairly easily, e.g.,
  by viscosity or magnetic fields
  \citep[][]{Thompson_Quataert_Burrows__2004__ApJ__Vis_Rot_SN}.
  Additionally, a differentially rotating magnetised fluid can be
  unstable against the magneto-rotational instability (MRI)
  \citep[][]{Balbus_Hawley__1991__ApJ__MRI,Balbus_Hawley__1998__RMP__MRI},
  leading to field amplification, turbulence and enhanced transport of
  angular momentum.
\end{description}

None of these mechanisms has been able to provide a robust explanation
for SNe across the entire mass range.  Because of the diversity and
complexity of the underlying physics, the problem is studied using
complementary approaches: very detailed simulations including the best
treatment of the microphysics that can be afforded and, on the other
hand, simplified models neglecting most of these complications for the
benefit of covering a larger parameter space of initial conditions.
The study of (magneto-)rotational mechanism relies currently mostly on
simplified models.

\cite{Akiyama_etal__2003__ApJ__MRI_SN} have pointed out that the most
basic instability condition of the MRI, i.e., a negative radial
gradient of the rotational profile, $\Omega (r)$, is typically
fulfilled in rotating post-collapse cores.  They have, based on
simplified spherical modelling, estimated growth times of a few
milliseconds and saturation fields of the order of $10^{15} ~
\mathrm{G}$.  Simplified global simulations
\citep[e.g.,][]{Obergaulinger_Aloy_Mueller__2006__AA__MR_collapse}
have shown that fields of this range (reached only for already
strongly magnetised progenitors with fields of the order $10^{12} ~
\mathrm{G}$) can have significant effects on the dynamics of the
explosion, e.g., by extracting rotational energy from the inner core
or launching jet-like outflows.  Additionally, signs for the growth of
the MRI were found in a few of those simulations.  Based only on such
simplified studies, a thorough assessment of the influence of the MRI
is still not possible.

\section{Methods and Aim of our Study}
\label{Sek:Mth}

The main obstacle for investigating the MRI by detailed microphysical
simulations is the necessity to resolve the wavelength of the most
unstable MRI modes, $\lambda_{\mathrm{MRI}}$, which scales with the
local Alfv\'en velocity, leading to a prohibitive resolution
requirement: for a realistic initial field of $10^{9} ~ \mathrm{G}$, a
region of several 100 kilometres would have to be covered by a 3d grid
with a resolution of a few centimetres or metres.  Consequently, we
investigated the properties of the MRI in core-collapse SNe using
high-resolution simulations of small representative domains in the
unstable region
\citep[][]{Obergaulinger_etal__2009__AA__Semi-global_MRI_CCSN}.

We studied basic properties of the MRI in the linear regime under
typical conditions of post-collapse cores differing from the
well-studied case of accretion discs mainly by the spherical geometry
and the potential importance of thermal stratification exerting either
a stabilising or a destabilising effect on the MRI.  Since the
absolute value of the Brunt-V{\"a}is{\"a}l{\"a} frequency, $N$, of the
star can exceed $\Omega$ strongly, the dynamics of the core may not be
dominated by standard MRI modes but rather show a more complex
interplay between the MRI and hydromagnetic convection.

Even if the instability grows fast enough, it can affect the explosion
only if a sufficiently large mean total (Maxwell plus Reynolds) stress
tensor is reached at the termination of the growth and maintained
during the turbulent saturated phase.  This is a long-standing issue
in MRI theory, and no final solution has been found although
simulations have identified a number of factors, physical and
numerical, affecting the amplitudes.  We tried to estimate the
saturation amplitude by an analysis of the non-linear phase of our
models.

\section{Results}
\label{Sek:Res}

\subsection{Instability Analysis}
\label{sSek:Insan}

We analysed the dispersion relation of wave-like perturbations of a
background in rotational equilibrium endowed with a weak magnetic
field with given rotational profile and Brunt-V{\"a}is{\"a}l{\"a}
frequency, $N$, i.e., the oscillation frequency of a buoyant fluid
element (which is imaginary in a convectively unstable layer).
Different from accretion discs, the main stabilising agent
counteracting gravity is not rotation but pressure.  Thus, the
corresponding models possess a sub-Keplerian rotation.

The instability criteria, given by
\cite{Balbus__1995__ApJ__stratified_MRI}, allow for two branches of
unstable modes: \emph{buoyant} and \emph{Alfv\'en} modes,
corresponding to unstable buoyant oscillations and unstable Alfv\'en
waves, respectively.  Both branches are stabilised by magnetic tension
at very short wavelengths.

Apart from a stable region of strongly positive gradients of entropy
or $\Omega$, these modes dominate in two different regions of the
parameter space.  For a negative entropy gradient and moderate
(positive or negative) values of $N^2$, corresponding to a small
influence of the thermal stratification, Alfv\'en modes dominate.  They
grow most rapidly in a narrow wavelength region around
$\lambda_{\mathrm{MRI}}$ and have a vanishing growth rate at infinite
wavelength, leading to stringent resolution requirements in numerical
simulations.  If $- N^2 > 0$ exceeds a threshold, buoyant modes set
in, while Alfv\'en modes disappear.  Similar to convection in a
non-magnetised fluid, they grow rapidly from infinite to short
wavelengths.

Based on the relative importance of these modes, we can distinguish
six regions in the space given by the Brunt-V{\"a}is{\"a}l{\"a}
frequency and the rotational profile: a stable region, the classic
(accretion-disc) MRI, Rayleigh instability, magnetoconvection, a
region in which the stabilisation of convection due to rapid (rigid)
rotation is lifted by magnetic fields, and a mixed regime,
respectively.

\subsection{Numerical Models}
\label{sSek:Nummo}

We performed a large parameter study of simplified 2d and 3d
simulations representing small (a few kilometres wide) sections of the
equatorial region of a post-collapse core with a cylindrical mesh of a
resolution between 0.625 and 20 metres.  We added a weak magnetic
field in vertical, $z$, direction, either uniform or sinusoidally
varying with a vanishing net flux, to the gas assumed in rotational
equilibrium with a prescribed rotational profile and entropy gradient.

To save computational costs, we neglected neutrino transport entirely
and replaced the complex nuclear equation of state by a simple
analytic approximation.  Our numerical code is based on a conservative
formulation of the equations of ideal MHD within the
constrained-transport framework
\citep[][]{Evans_Hawley__1998__ApJ__CTM} and uses high-order methods
for reconstruction \citep[e.g., the monotonicity-preserving schemes
by][]{Suresh_Huynh__1997__JCP__MP-schemes} and approximate Riemann
solvers.

\begin{figure}[!ht]
  \label{Fig:MRI-3dstruct}
  \centering
   \includegraphics[width=4.3cm]{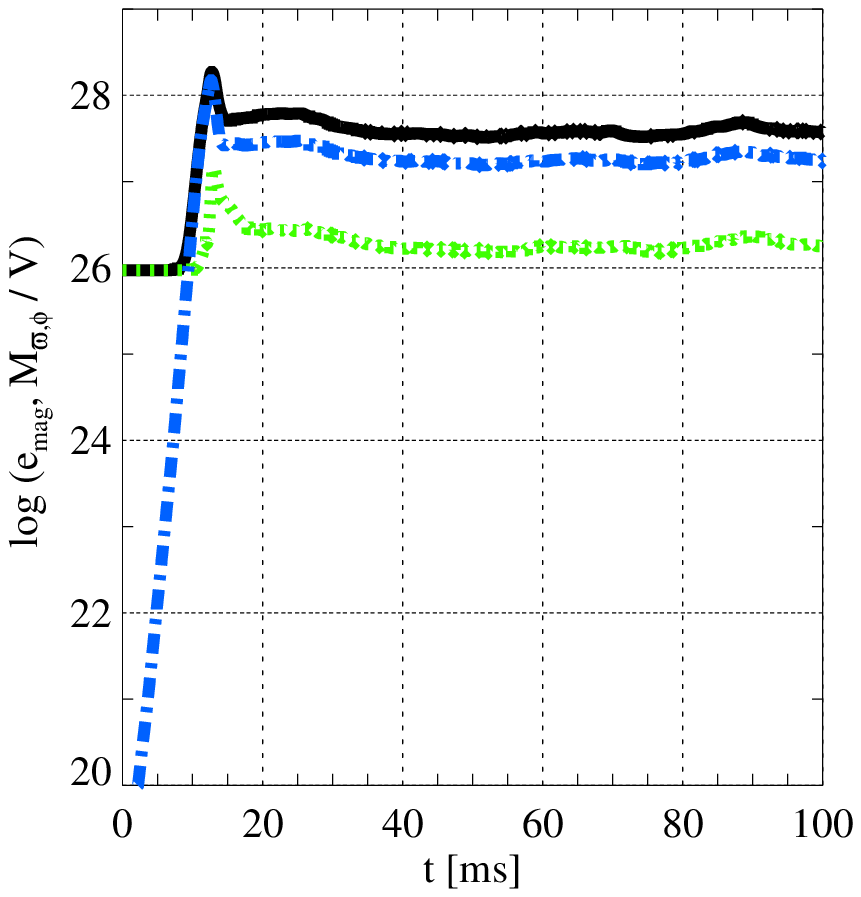}
   \includegraphics[width=6.6cm]{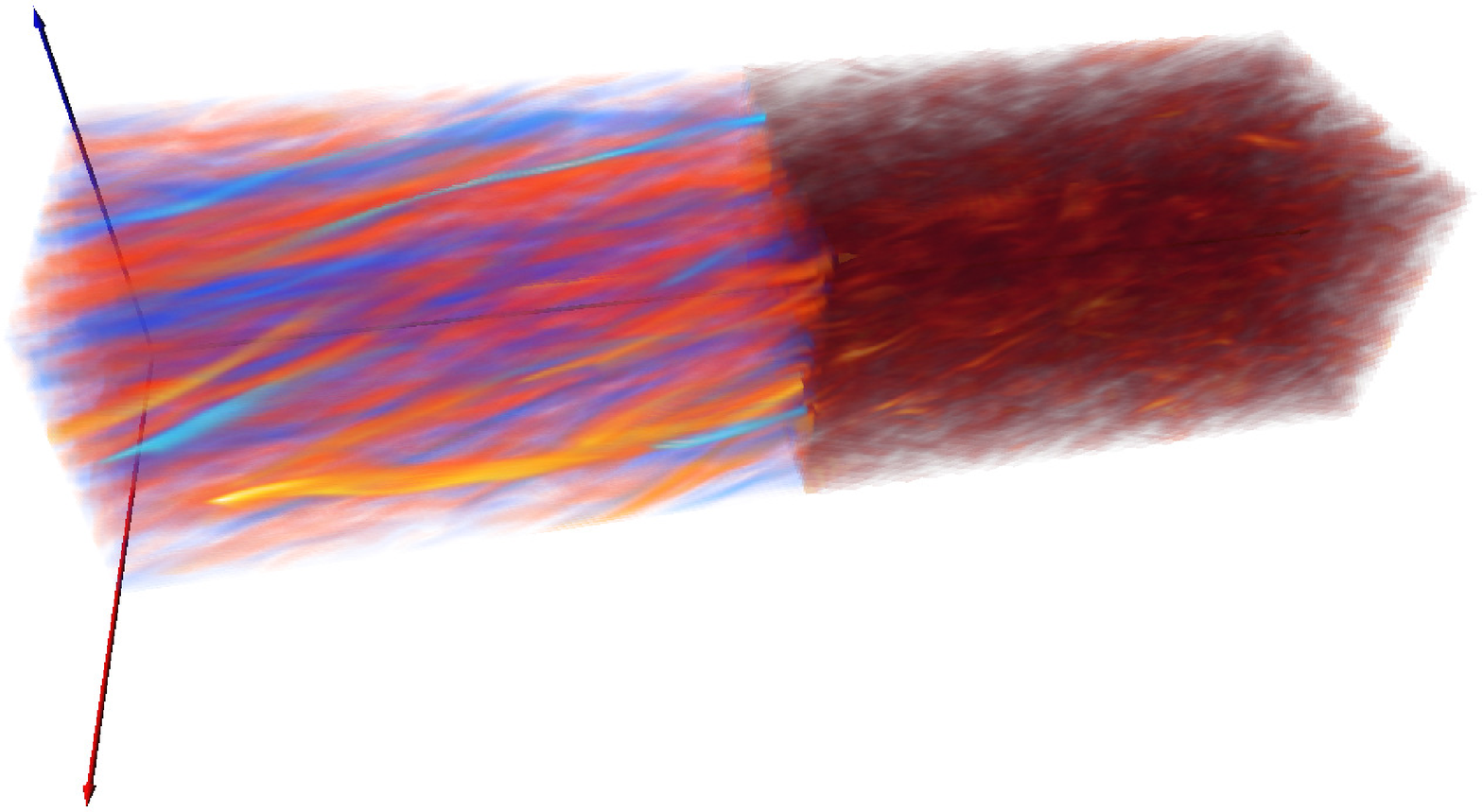}
  \caption{ The time evolution of a model with vanishing net flux of
    the magnetic field (left panel) and its spatial structure in the
    saturated state (right panel).  The left panel shows the total
    magnetic energy (black solid line), its $z$-component (green
    dotted line), and the Maxwell stress component $M_{\varpi\phi} =
    b_{\varpi} b_{\phi}$ (blue dash-dotted line) per unit volume as a
    function of time in milliseconds.  The right panel shows the
    computational box; the radial, azimuthal, and vertical directions
    are indicated by the red (pointing towards the observer), green
    (hidden, pointing to the right), and blue (pointing up) arrows.
    The left and right halves of the figure show a volume rendering of
    the toroidal magnetic field, $b_{\phi}$, (positive and negative
    values are shown by blue and red colours) and of the vorticity of
    the flow, respectively.  }
\end{figure}

The simulations confirm all regimes relevant to the MRI in SNe, with
numerical growth rates very well in agreement with the analytic
predictions.  Under conditions of a rapidly rotating core, we expect
typical MRI growth times of the order of a few milliseconds, i.e.,
sufficiently fast to affect the explosion developing on time scales of
several tens or hundreds of milliseconds (see the left panel of
Fig.\,\ref{Fig:MRI-3dstruct} for the time evolution of a model with
vanishing net flux).

The initial perturbations grow exponentially at the MRI growth rate
given by linear analysis, $\sigma_{\mathrm{MRI}}$.  Their growth is
mediated by coherent large-scale \emph{channel modes}, i.e.,
axisymmetric pairs of upflows and downflows with alternating magnetic
polarity.  As shown by \citet{Goodman_Xu__1994__ApJ__MRI-parasitic},
the channel modes are themselves unstable against secondary,
\emph{parasitic} instabilities of Kelvin-Helmholtz or tearing-mode
type.  Our models show the growth of these parasites feeding off the
channel modes.  Their growth rate is a function of the field strength
of the channels.  The higher the field strength of a channel is, the
faster grow its parasites, until, at a certain field strength, the
amplitude of the parasites exceeds that of the channel: the MRI growth
terminates.  In the left panel of Fig.\,\ref{Fig:MRI-3dstruct}, the
growth of these parasites can be identified by the strong increase of
the $z$-component of the magnetic field near $t = 12 ~ \mathrm{ms}$.
Typically, the parasitic mode quenching the MRI in our models is of
tearing-mode type.  This is clearly an artefact of our numerics.
Since we evolve the system of \emph{ideal} MHD there should be no
resistivity, and thus no tearing-mode instability.  However, the
unavoidable numerical resistivity, albeit only at a small level,
permits the growth of resistive instabilities.

Combining the MRI growth rates from linear analysis and those of the
parasites determined in a series of auxiliary simulations, we
establish scaling laws for the value of the Maxwell stress at the
termination of MRI growth.  A stronger initial field, $b_0$,
corresponds to wider channel flows which can sustain a stronger field
before they are disrupted by parasites.  Consequently, the termination
amplitude scales with the initial field as $(b_0)^{16/7}$.  A higher
grid resolution, $\delta$, corresponds to a smaller value of the
numerical resistivity, and thus to a slower growth of the parasites,
leading to a scaling of the termination amplitude as $\delta^{-8/7}$.
Slower rotation leads to wider channels, again decreasing the growth
of parasites.  Hence we find a scaling with $\Omega^{8/7}$.  Finally,
the growth rate of the parasites depends on the local sound speed,
yielding a scaling with $c_{\mathrm{s}}^{6/7}$.

After the disruption of the coherent channel flows, turbulence
develops (see Fig.\,\ref{Fig:MRI-3dstruct}, right panel, for the
structure of a model in this state).  The flow and the magnetic field
(shown in the right and left halves of the panel) are characterised by
a multitude of small-scale filamentary features.  Most obvious are
magnetic flux tubes elongated in azimuthal direction.  Depending on
the MRI regime and the predominance of Alfv\'en (buoyant) modes,
efficient transport of angular momentum (entropy) leads to rigid
rotation (flat entropy profile, and a rotation profile where the
specific angular momentum is constant).

The Maxwell stress is typically maintained at a level of the order of
its termination value throughout saturation, but it is subject to
pronounced oscillations and even intermittent episodes of exponential
growth.  These correspond to the re-appearance of coherent
channel-like flows that terminate after a short time by parasitic modes
\citep[][]{Sano_Inutsuka__2001__ApJL__MRI-recurrent-channels}.  Apart
from these flows, we identify large-scale correlations of the velocity
and magnetic field in the turbulence maintained over several orbital
periods, similar to the ones observed by
\citet{Lesur_Ogilvie__2008__AA__zero_flux_MRI_dynamo}.

Our results suggest that the MRI may play a role, for SNe that have
rapidly rotating progenitors.  The field strengths reached in our
models are of the order of $10^{15} ~ \mathrm{G}$, i.e., sufficiently
strong to affect the explosion by redistributing angular momentum and
extracting energy from rotation.  For slower rotation, the interplay
of the MRI and convection, i.e., the mixed instability regime may play
a role in the unstable regions of the PNS and the surrounding
neutrino-heated hot bubble.  Neglecting a large part of the relevant
physics and restricting ourselves to a rather special geometry, we
are, however, far from a final answer, e.g., in the form of a MRI
model with predictive power for generic post-collapse cores.  Further
modelling is required to include the effects neglected thus far.

\acknowledgements This research has been supported by the Spanish {\it
  Ministerio de Educaci\'on y Ciencia} (grants AYA2007-67626-C03-01,
CSD2007-00050), and by the Collaborative Research Center on {\it
  Gravitational Wave Astronomy} of the Deutsche Forschungsgemeinschaft
(DFG SFB/Transregio 7).  MAA is a Ram\'on y Cajal fellow of the {\em
  Ministerio de Educaci\'on y Ciencia}.  Most of the simulations were
performed at the Rechenzentrum Garching (RZG) of the
Max-Planck-Society. We are also thankful for the computer resources,
the technical expertise, and the assistance provided by the Barcelona
Supercomputing Center - Centro Nacional de
Supercomputaci\'on. Finally, we would like to A.~Marek for helpful
discussions.

\end{document}